# Superconductivity and Intercalation State in the Lithium-Hexamethylenediamine-Intercalated Superconductor Li$_x$(C$_6$H$_{16}$N$_2$)$_y$Fe$_{2-z}$Se$_2$ : Dependence on the Intercalation Temperature and Lithium Content

Shohei Hosono, Takashi Noji, Takehiro Hatakeda, Takayuki Kawamata, Masatsune Kato, and Yoji Koike

*Department of Applied Physics, Tohoku University, 6-6-05 Aoba, Aramaki, Aoba-ku, Sendai 980-8579, Japan*

The superconductivity and intercalation state in the lithium- and hexamethylenediamine (HMDA)-intercalated superconductor Li$_x$(C$_6$H$_{16}$N$_2$)$_y$Fe$_{2-z}$Se$_2$ have been investigated from powder x-ray diffraction, thermogravimetric and magnetic susceptibility measurements, changing the intercalation temperature, $T_i$, and the Li content, $x$. Both Li and HMDA have been co-intercalated stably up to $x = 2$ roughly in the molar ratio of $x : y = 2 : 1$. In the case of $T_i = 45$ ˚C, it has been found that both Li and HMDA are co-intercalated locally at the edge of FeSe crystals, indicating that both Li and HMDA are hard to diffuse into the inside of FeSe crystals at 45 ˚C. In the case of $T_i = 100$ ˚C, on the other hand, it has been found that both Li and HMDA diffuse into the inside of FeSe crystals, so that $T_c$ tends to increase with increasing $x$ from ~ 30 K at $x = 1$ up to 38 K at $x = 2$ owing to the increase of electron carriers doped from Li into the FeSe layers.

Keywords : Superconductivity, FeSe, Intercalation, Lithium, Hexamethylenediamine,

# 1. Introduction

The superconducting transition temperature, $T_c$, of the iron-based layered superconductor FeSe is as low as 8 K,[1] but it markedly increases through the co-intercalation of alkali or alkaline-earth metals and ammonia or organic molecules between FeSe layers. Collecting the data of $T_c$ and the interlayer spacing between neighboring Fe layers, $d$, of a variety of FeSe-based intercalation superconductors,[2-26] it has been found that $T_c$ tends to increase with increasing $d$ and to be saturated at ~ 45 K for $d \gtrsim 9$ Å.[12] In single-layer FeSe films, on the other hand, the opening of a possible superconducting gap has been observed at low temperatures below 42 - 65 K in the scanning tunneling microscopy/spectroscopy[27] and angle-resolved photoemission spectroscopy[28,29] measurements. Since the $d$ value of single-layer FeSe films may be regarded as infinite, $T_c$ is expected to increase with further increasing $d$ for FeSe-based intercalation superconductors.

Thus $T_c$'s of the FeSe-based intercalation superconductors are dominated by the $d$ values, while the electron doping from intercalants into the conducting FeSe layers seems not to affect $T_c$ so much.[12] These results have been interpreted by Guterding et al.[30] in terms of the electronic structure calculated with the density functional theory. That is, the electronic structure becomes more two-dimensional with increasing $d$, leading to the increase in $T_c$, and it is perfectly two-dimensional at $d = 8 - 10$ Å, leading to the saturation of $T_c$ at $d \geqq 8$ Å. Within the random-phase-approximation spin-fluctuation approach, moreover, they have predicted that $T_c$'s of the FeSe-based intercalation superconductors with the $S_\pm$ symmetry increase not through the further increase in $d$ but through the electron doping owing to the increase of the density of states at the Fermi level at $d \geqq 8$ Å. In linear-diamine-intercalated $A_x(C_nH_{2n+4}N_2)_yFe_{2-z}Se_2$ ($A$ = Na, Sr ; $n$ = 0, 2, 3) with $d$ = 8.7 - 11.4 Å, in fact, $T_c$ is not so dependent on $d$ and $T_c$ = 41 – 46 K of heavily electron-doped Na-intercalated compounds are higher than $T_c$ = 34 – 38 K of lightly electron-doped Sr-intercalated ones.[18] Also in the lithium- and hexamethylenediamine (HMDA)-intercalated superconductor $Li_x(C_6H_{16}N_2)_yFe_{2-z}Se_2$, which we have recently discovered, $T_c$ = 41 K of the more electron-doped post-annealed sample with $d$ ~ 10 Å is higher than $T_c$ = 38 K of the less electron-doped as-intercalated sample with $d$ ~ 16 Å, which is the largest among those of the FeSe-based intercalation compounds.[13,26] Thus the difference in $T_c$ at $d \geqq 8$ Å is roughly explained according to their prediction. However, the detailed electron-doping dependence of $T_c$ in the FeSe-based intercalation superconductors has not yet been clarified. Moreover, the intercalation state in these superconductors has not understood well.

In the present work, we have synthesized lithium- and HMDA-intercalated $Li_x(C_6H_{16}N_2)_yFe_{2-z}Se_2$ with various values of the Li content, $x$, to investigate the electron-doping dependence of $T_c$. Changing the intercalation temperature, $T_i$, we have also studied the intercalation state from powder x-ray diffraction, thermogravimetric (TG) and magnetic susceptibility, $\chi$, measurements.

## 2. Experimental

Polycrystalline host samples of FeSe were prepared by the high-temperature solid-state reaction method as described in Ref. 11. Iron powder and selenium grains, which were weighted in the molar ratio of Fe : Se = 1.02 : 1, were mixed, put into an alumina crucible, and sealed in an evacuated quartz tube. This was heated at 1027 ˚C for 30 h and then annealed at 400 ˚C for 50 h, followed by furnace-cooling. The obtained ingot of FeSe was pulverized into powder to be used for the intercalation. A series of samples of $Li_x(C_6H_{16}N_2)_yFe_{2-z}Se_2$ with various $x$ values were prepared according to the method described in our previous papers.[13,26] First, an appropriate amount of powdery FeSe was placed in a beaker filled with solution of pure Li dissolved in HMDA. Assuming the whole Li in the solution to be intercalated into FeSe, the amount of Li was set as $0.25 \leqq x \leqq 4$ in the molar ratio of Li : FeSe = $x$ : 2. The reaction of intercalation was carried out at $T_i$ = 45 ˚C or 100 ˚C for 3 days. After the reaction, the separation of the product from residual HMDA was easily performed by the solidification of residual HMDA at the top cap of the beaker, keeping the temperature of the top cap of the beaker below the melting point of HMDA (42 ˚C). All the processes were carried out in an argon-filled glove box.

Both the host sample of FeSe and as-intercalated samples covered with an airtight sample-holder were characterized by the powder x-ray diffraction using $CuK_\alpha$ radiation. The diffraction patterns were analyzed using RIETAN-FP.[31] TG measurements were performed in flowing gas of argon, using a commercial analyzer (SII Nano Technology Inc., TG/DTA7300). To detect the superconducting transition, $\chi$ was measured using a superconducting quantum interference device (SQUID) magnetometer (Quantum Design, MPMS).

## 3. Results and Discussion

Figure 1(a) shows powder x-ray diffraction patterns of the host sample of FeSe and as-intercalated samples with various nominal values of $x$ synthesized at 45 ˚C. As reported in our previous papers,[13,26] it is found that the sample of $x$ = 1 is mainly composed of Phase I of $Li_x(C_6H_{16}N_2)_yFe_{2-z}Se_2$ with $d \sim 16$ Å, as shown in Fig. 2(a), and that there remain small regions of non-intercalated FeSe in the sample.

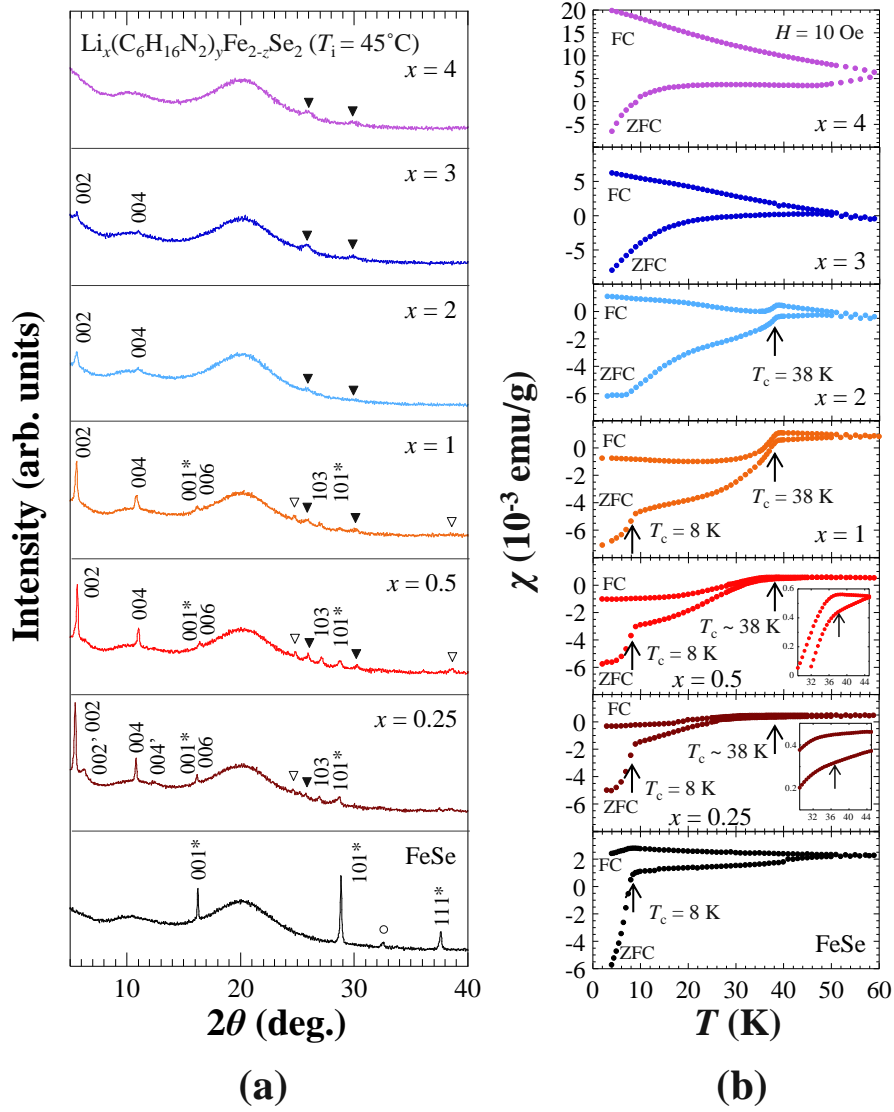

Fig. 1. (a) Powder x-ray diffraction patterns of the host sample of FeSe and as-intercalated samples with various nominal values of $x$ synthesized at 45 °C consisting of $Li_x(C_6H_{16}N_2)_yFe_{2-z}Se_2$ (Phase I and phase II) and FeSe, using CuK$_\alpha$ radiation. Indices without and with an asterisk are based on the ThCr$_2$Si$_2$-type and PbO-type structures, respectively. Indices of the ThCr$_2$Si$_2$-type structure without and with a single quotation mark represent phase I and Phase II, respectively. Peaks marked by ▼, ○, and ▽ are due to Li$_2$Se, Fe$_7$Se$_8$, and an unknown compound, respectively. The broad peak around $2\theta = 20°$ is due to the airtight sample-holder. (b) Temperature dependence of the magnetic susceptibility, $\chi$, in a magnetic field of 10 Oe on zero-field cooling (ZFC) and field cooling (FC) for the host sample of FeSe and as-intercalated samples with various nominal values of $x$ synthesized at 45 °C consisting of $Li_x(C_6H_{16}N_2)_yFe_{2-z}Se_2$ and FeSe. Insets of x = 0.25 and 0.5 show the temperature dependence of $\chi$ around $T_c$.

Similarly, Bragg peaks due to Phase I are predominantly observed in the samples of $0.25 \leqq x \leqq 3$, indicating that the obtained main phase is not dependent on $x$. However, it is found that the intensity of Bragg peaks due to Phase I decreases with increasing $x$ and that no Bragg peaks due to Phase I is observed in the sample of $x = 4$. As for the sample of $x = 0.25$, it is found that there are small regions of Phase II, where intercalated molecules of HMDA are inclined ~ 60° from the FeSe layers and the $d$ value of ~ 14 Å is slightly shorter than $d$ ~ 16 Å of Phase I, as shown in Fig. 2(b).[26] Both the lattice constant $c$ and $d$ of $Li_x(C_6H_{16}N_2)_yFe_{2-z}Se_2$ are listed in Table I.

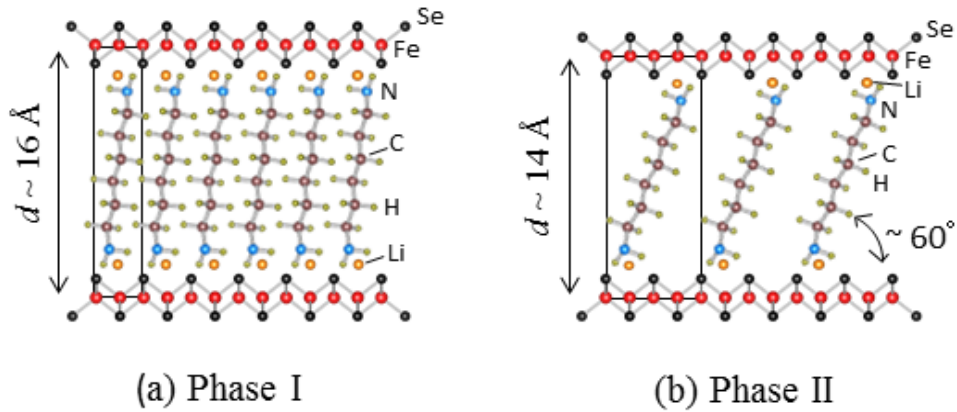

Fig. 2. Schematic views of crystal structures of (a) Phase I of $Li_x(C_6H_{16}N_2)_yFe_{2-z}Se_2$ with $d$ ~ 16 Å, and (b) Phase II of $Li_x(C_6H_{16}N_2)_yFe_{2-z}Se_2$ with $d$ ~ 14 Å. Solid lines represent halves of the respective unit-cells.

Figure 3(a) shows TG curves on heating up to 900 °C at the rate of 1 °C/min for as-intercalated samples with various nominal values of $x$ synthesized at 45 °C. As reported in our previous paper,[13] the mass loss below ~ 250 °C is in good correspondence with the weight percent of intercalated HMDA estimated by the inductively coupled plasma optical emission spectroscopy for the sample of $x = 1$. Therefore, the content of HMDA, $y$, can be approximately estimated from the mass loss below 250 °C, as listed in Table II and shown in Fig. 4. It is found that the $y$ value increases with increasing $x$ and is roughly a half of the $x$ value for $x \leqq 2$. This is reasonable, because $Li^+$ ions seem to be located near lone-pair electrons of N atms at both edges of linear molecules of HMDA, as shown in Fig. 2. Here, it is noted that the observation of Phase II only in the sample of $x = 0.25$ with a small $y$ value, as shown in Fig. 1(a), is reasonable, because Phase II is stabilized in a region where HMDA molecules are not packed densely, as shown in Fig. 2(b). As for the samples of $x = 3$ and 4, the $y$ value is found to deviate from the linear relation to the $x$ value. Taking into account the observation of no Bragg peaks due to Phase I in the sample of $x = 4$ as mentioned above, this will be attributed to the decomposition of the FeSe layers owing to the strong reaction between Li and FeSe.

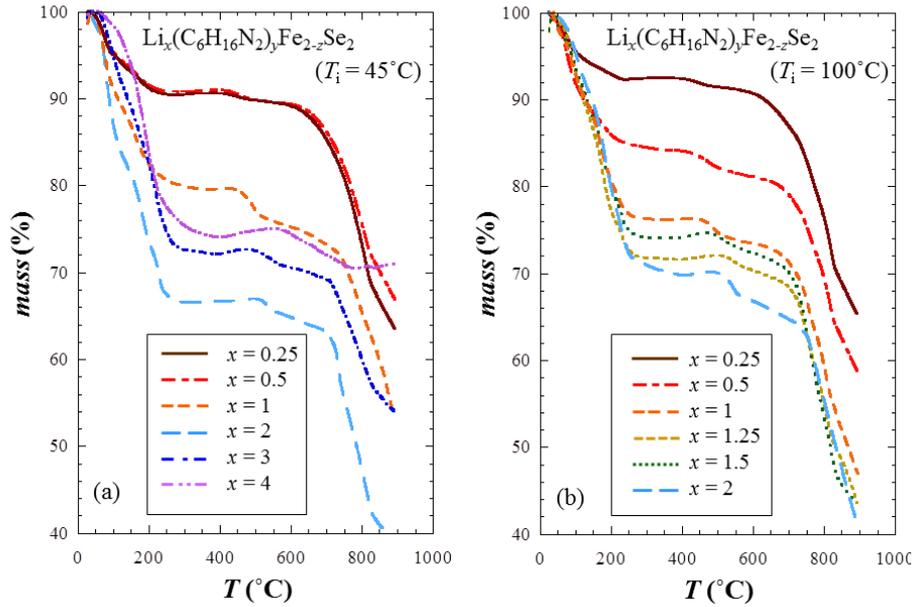

Fig. 3. Thermogravimetric curves on heating at the rate of 1 ˚C/min for as-intercalated samples with various nominal values of $x$ synthesized (a) at 45 ˚C and (b) at 100 ˚C.

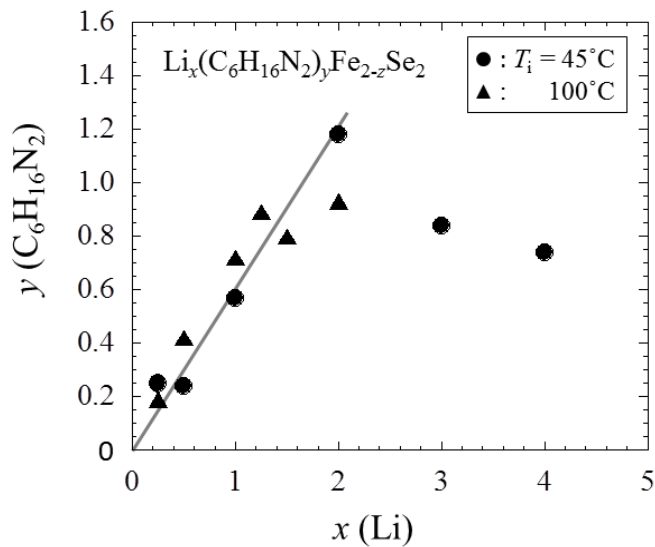

Fig. 4. Dependence on the nominal Li content, $x$, of the content of $C_6H_{16}N_2$, $y$, in $Li_x(C_6H_{16}N_2)_yFe_{2-z}Se_2$ estimated from the mass loss below 250 ˚C in the thermogravimetric curve for as-intercalated samples with various nominal values of $x$, as shown in Figs. 3(a) and 3(b).

Figure 1(b) shows the temperature dependence of $\chi$ in a magnetic field of 10 Oe on zero-field cooling (ZFC) and on field cooling (FC) for the host sample of FeSe and as-intercalated samples with various nominal values of $x$ synthesized at 45 °C. As reported in our previous papers,[13,26] the first and second transitions at 38 K and 8 K in the sample of $x = 1$ are due to bulk superconductivity of Phase I and non-intercalated FeSe, respectively. Likewise, the superconducting transition at ~ 38 K is observed in the samples of $0.25 \leqq x \leqq 2$, though $T_c$'s of x = 0.25 and 0.5 are defined at the onset temperature of the superconducting transition, namely, the temperature where $\chi$ deviates from the linear temperature dependence at high temperatures because of the rather broad transitions and therefore have large ambiguity. However, it is found that the superconducting volume fraction of Phase I, roughly estimated from the change in $\chi$ on ZFC between 38 K and 8 K, increases with increasing $x$, while the superconducting volume fraction of non-intercalated FeSe, roughly estimated from the change in $\chi$ on ZFC between 8 K and the measured lowest temperature, decreases with increasing $x$. Taking into account the powder x-ray diffraction result indicating the coexistence of Phase I and non-intercalated FeSe, as shown in Fig. 1(a), it appears that both Li and HMDA are only co-intercalated locally at the edge of FeSe crystals and are hard to diffuse into the inside of FeSe crystals in the case of $T_i = 45$ °C. In this case, it is understood that both Li$^+$ ions and HMDA molecules are densely packed at the edge of FeSe crystals, so that the electron-doping level in the co-intercalated regions of Phase I does not change, leading to the almost same $T_c$ ~ 38 K for $0.25 \leqq x \leqq 2$. As for no superconducting transition in the samples of $x = 3$ and 4, it will be due to the decomposition of the FeSe layers owing to the strong reaction with Li, as mentioned above.

Figure 5(a) shows powder x-ray diffraction patterns of FeSe and as-intercalated samples with various nominal values of $x$ synthesized at 100 °C. The samples of $0.25 \leqq x \leqq 3$ are mainly composed of Phase I and Phase II of Li$_x$(C$_6$H$_{16}$N$_2$)$_y$Fe$_{2-z}$Se$_2$ and that there remain very small regions of non-intercalated FeSe in these samples. It is found that Phase II is dominant in the sample of $x = 0.25$, while Phase I is dominant for $0.5 \leqq x \leqq 3$. Both the lattice constant $c$ and $d$ of Li$_x$(C$_6$H$_{16}$N$_2$)$_y$Fe$_{2-z}$Se$_2$ are listed in Table I.

Figure 3(b) shows TG curves on heating up to 900 °C at the rate of 1 °C/min for as-intercalated samples with various nominal values of $x$ synthesized at 100 °C. The TG curves are analogous to those of the as-intercalated samples synthesized at 45 °C, as shown in Fig. 3(a). As listed in Table II and shown in Fig. 4, the $x$ dependence of the $y$ value estimated from the mass loss below 250 °C in the samples synthesized at 100 ˚C is similar to that of the samples synthesized at 45 ˚C, indicating that one molecule of HMDA is approximately intercalated together with two Li$^+$ ions irrespective of $T_i$. Accordingly, it is reasonable that Phase II, which is stabilized in a region where HMDA molecules are not packed so densely, is dominant in the sample of $x = 0.25$ with a small $y$ value synthesized at 100 ˚C.

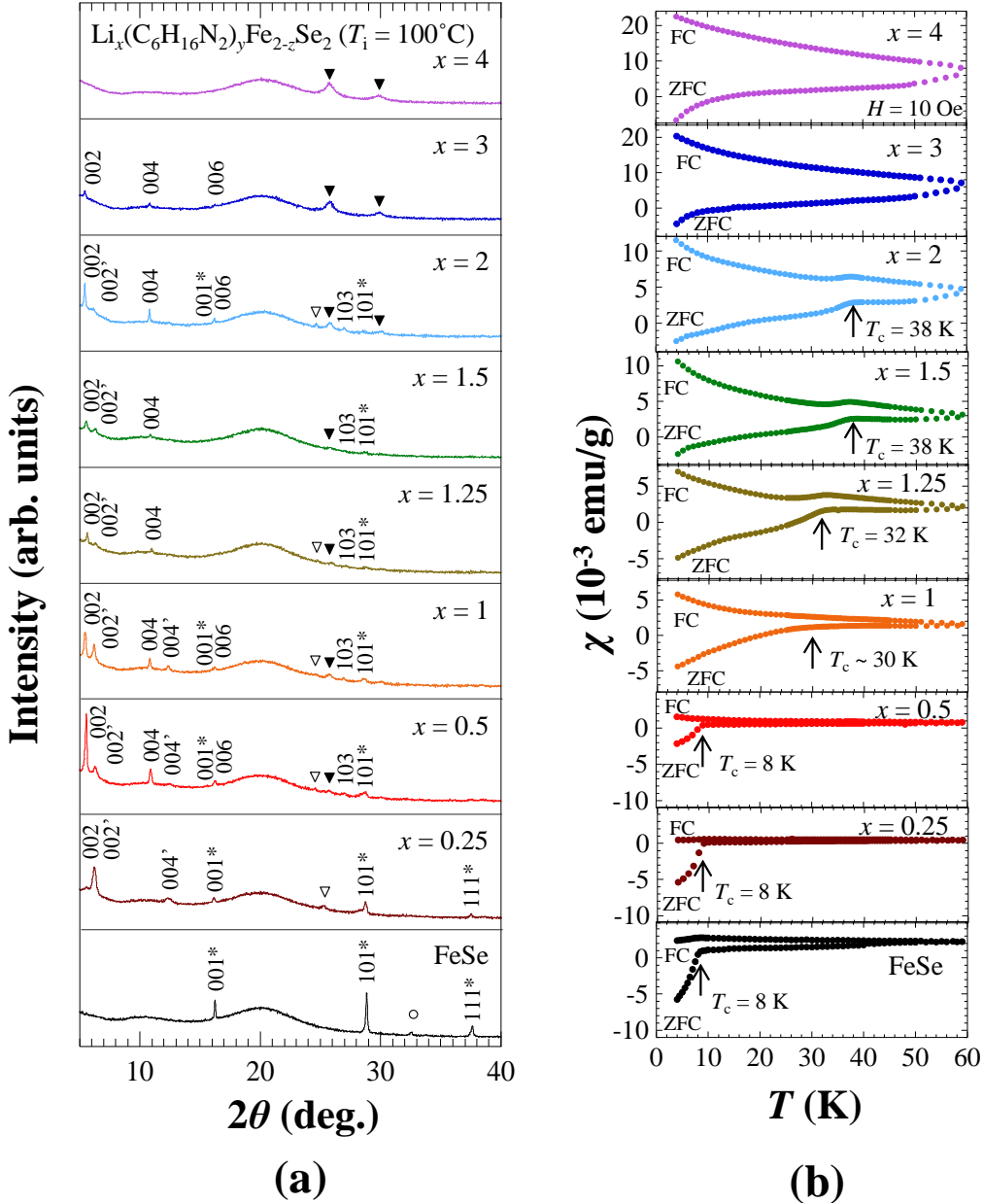

Fig. 5. (a) Powder x-ray diffraction patterns of the host sample of FeSe and as-intercalated samples with various nominal values of $x$ synthesized at 100 ˚C consisting of Li$_x$(C$_6$H$_{16}$N$_2$)$_y$Fe$_{2-z}$Se$_2$ (Phase I and Phase II) and FeSe, using CuK$_\alpha$ radiation. Indices without and with an asterisk are based on the ThCr$_2$Si$_2$-type and PbO-type structures, respectively. Indices of the ThCr$_2$Si$_2$-type structure without and with a single quotation mark represent Phase I and Phase II, respectively. Peaks marked by ▼, ○, and ▽ are due to Li$_2$Se, Fe$_7$Se$_8$, and unknown compounds, respectively. The broad peak around $2\theta = 20°$ is due to the airtight sample-holder. (b) Temperature dependence of the magnetic susceptibility, $\chi$, in a magnetic field of 10 Oe on zero-field cooling (ZFC) and field cooling (FC) for the host sample of FeSe and as-intercalated samples with various nominal values of $x$ synthesized at 100 °C consisting of $Li_x(C_6H_{16}N_2)_yFe_{2-z}Se_2$ and FeSe.

Figure 5(b) shows the temperature dependence of $\chi$ in a magnetic field of 10 Oe on ZFC and on FC for FeSe and as-intercalated samples with various nominal values of $x$ synthesized at 100 °C. Unlike the samples synthesized at 45 °C, a single superconducting transition is observed at ~ 30 K in the sample of $x = 1$, though the $T_c$ is defined at the onset temperature of the superconducting transition because of the rather broad transition. It is found that $T_c$ of $Li_x(C_6H_{16}N_2)_yFe_{2-z}Se_2$ tends to increase with increasing $x$ in the samples of $1 \leqq x \leqq 2$. In the samples of $x = 0.25$ and 0.5, only the superconducting transition of non-intercalated FeSe is observed, while no superconducting transition due to $Li_x(C_6H_{16}N_2)_yFe_{2-z}Se_2$ is observed in spite of the clear observation of Bragg peaks due to Phase II and Phase I, respectively, as shown in Fig. 5(a), though the reason is not clear. As for the samples of $x = 3$ and 4, no superconducting transition will be due to the decomposition of the FeSe layers as in the case of the samples synthesized at 45 °C. Here, it is noted that the value of $\chi$ in the normal state tends to increase with increasing $x$ and that the hysteresis of $\chi$ in the normal state between ZFC and FC becomes marked with increasing $x$. These behaviors are observed also in the samples synthesized at 45 °C, as shown in Fig. 1(b), and are inferred to be due to magnetic impurities such as Fe produced by the reaction between Li and FeSe. The hysteresis of $\chi$ in the normal state of the host sample of FeSe is probably due to magnetic impurities such as $Fe_7Se_8$.

Finally, taking into account the powder x-ray diffraction and magnetic susceptibility results of the as-intercalated samples synthesized at 45 °C and 100 °C, we discuss the intercalation state. As shown in Fig. 6(a), in the sample of $x = 1$ synthesized at 45 °C, two transitions at 38 K due to bulk superconductivity of Phase I and at 8 K due to non-intercalated FeSe have been observed, indicating that both Li and HMDA are only co-intercalated locally at the edge of FeSe crystals so that the sample is inhomogeneous. However, in the sample of $x = 1$ synthesized at 100 °C, a single superconducting transition has been detected at ~ 30 K and there has been no transition at 8 K due to FeSe, as shown in Fig. 6(b). The disappearance of the transition due to FeSe is attributed to the successful diffusion of intercalated Li into the inside of FeSe crystals, leading to the formation of a mixture of Phase I and Phase II. In the sample of $x = 2$ synthesized at 100 °C, a single superconducting transition due to Phase I has been observed, indicating that both Li and HMDA are co-intercalated rather homogeneously in this sample, as shown in Fig. 6(c). Accordingly, to synthesize a homogeneous sample of Phase I, it turns out that $T_i$ = 45 °C is too low and that 3 and more days are necessary at $T_i$ = 100 °C.   As shown in Fig. 6, moreover, the amount of Li per FeSe is found to be smaller in the sample of $x$ = 1 synthesized at 100 °C than at the edge of FeSe crystals in the sample of $x$ = 1 synthesized at 45 °C and also than in the sample of $x$ = 2 synthesized at 100 °C. Therefore, according to the prediction by Guterding et al.,[30] the low $T_c$ ~ 30 K in the sample of $x$ = 1 synthesized at 100 °C is understood to be due to the low electron-doping level, compared with the edge of FeSe crystals in the sample of $x$ = 1 synthesized at 45 °C and also the sample of $x$ = 2 synthesized at 100 °C.   Moreover, the increase in $T_c$ with increasing $x$ in the sample of $1 \leqq x \leqq 2$ synthesized at 100 °C is also understood to be due to the increase of electron carriers doped from Li into the FeSe layers.   Taking into account the decomposition of the FeSe layers in the samples of $x$ = 3 and 4, the maximum value of $x$ is 2 due to Phase I of $Li_x(C_6H_{16}N_2)_yFe_{2-z}Se_2$.   To increase $T_c$ further, therefore, more electron-doping from the other intercalants is necessary.

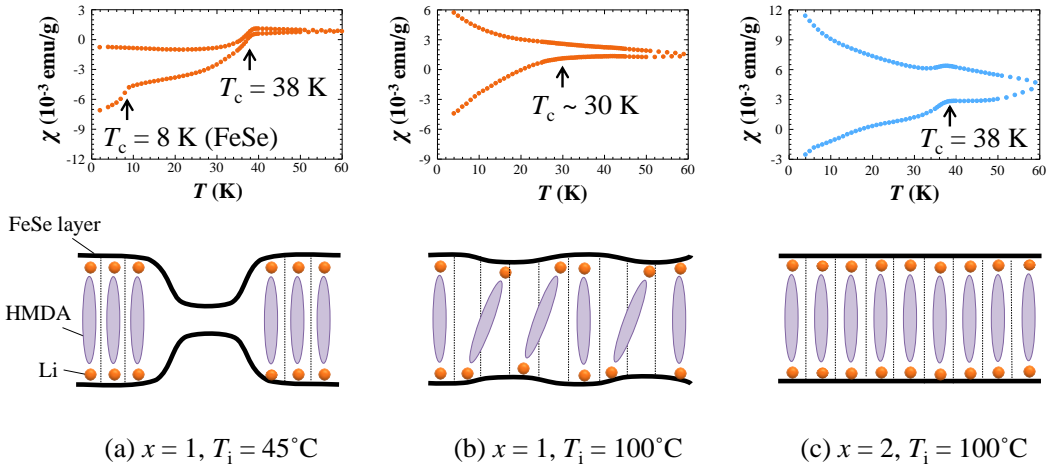

Fig. 6. Temperature dependence of the magnetic susceptibility, $\chi$, and the schematic view of the crystal structure for (a) the as-intercalated sample of $x$ = 1 synthesized at 45 °C, (b) the as-intercalated sample of $x$ = 1 synthesized at 100 °C, and (c) the as-intercalated sample of $x$ = 2 synthesized at 100 °C.

## 4. Summary

We have synthesized lithium- and HMDA-intercalated superconductor $Li_x(C_6H_{16}N_2)_yFe_{2-z}Se_2$, changing $T_i$ and $x$. Powder x-ray diffraction, TG and $\chi$ measurements have revealed that both Li and HMDA are co-intercalated stably up to $x$

= 2 roughly in the molar ratio of $x : y = 2 : 1$ and that the FeSe layers are decomposed for $x > 2$. In the case of $T_i = 45$ °C, two superconducting transitions at ~ 38 K due to Phase I of Li$_x$(C$_6$H$_{16}$N$_2$)$_y$Fe$_{2-z}$Se$_2$ with $x \sim 2$ and $d \sim 16$ Å and at 8 K due to non-intercalated FeSe have been observed in the samples of $0.25 \leqq x \leqq 2$, though the superconducting volume fraction of FeSe has decreased with increasing $x$. These results indicate that both Li and HMDA are only co-intercalated locally at the edge of FeSe crystals and are hard to diffuse into the inside of FeSe crystals at 45 °C. In the case of $T_i = 100$ °C, on the other hand, the samples of $1 \leqq x \leqq 2$ have mainly been composed of Phase I and Phase II and have shown a single superconducting transition without transition due to FeSe, indicating that both Li and HMDA diffuse into the inside of FeSe crystals. Moreover, $T_c$ has tended to increase with increasing $x$ from ~ 30 K at $x = 1$ up to 38 K at $x = 2$. The increase in $T_c$ has been understood to be due to the increase of electron carriers doped from Li into the FeSe layers, as predicted by Guterding et al.[21] To increase $T_c$ further, therefore, more electron-doping from the other intercalants is necessary.

**Acknowledgments**

This work was supported by JSPS KAKENHI Grant Number 15K13512.

Table I. Lattice constant $c$ (in Å) and the interlayer spacing between neighboring Fe layers, $d$ (in Å), of Li$_x$(C$_6$H$_{16}$N$_2$)$_y$Fe$_{2-z}$Se$_2$ with various nominal values of $x$ synthesized at the intercalation temperature, $T_i$, = 45 °C and 100 °C. It is noted that $d$ is given by the half of $c$.

| $T_i$ (°C) | Li content ($x$) | $c$ Phase I | $c$ Phase II | $d$ Phase I | $d$ Phase II |
|---|---|---|---|---|---|
| 45 | 0.25 | 32.81 (5) | 28.29 (1) | 16.40 (2) | 14.14 (1) |
| 45 | 0.5 | 32.30 (11) | | 16.15 (7) | |
| 45 | 1 | 32.73 (8) | | 16.37 (4) | |
| 45 | 2 | 32.40 (8) | | 16.20 (4) | |
| 45 | 3 | 32.06 (13) | | 16.03 (7) | |
| 100 | 0.25 | 32.70 | 28.258 (4) | 16.35 | 14.129 (2) |
| 100 | 0.5 | 32.58 (6) | 28.145 (2) | 16.29 (3) | 14.072 (1) |
| 100 | 1 | 32.77 (5) | 28.409 (6) | 16.39 (3) | 14.204 (3) |
| 100 | 1.25 | 31.91 (4) | 28.23 (2) | 15.96 (2) | 14.11 (1) |
| 100 | 1.5 | 32.44 (4) | 28.339 (9) | 16.22 (2) | 14.169 (5) |
| 100 | 2 | 32.77 (3) | 28.45 (4) | 16.39 (1) | 14.22 (2) |
| 100 | 3 | 32.67 (2) | | 16.34 (1) | |

Table II. Mass loss below 250 ˚C in the thermogravimetric curve on heating at the rate of 1˚C/min and the content of $C_6H_{16}N_2$, $y$, estimated from the mass loss for $Li_x(C_6H_{16}N_2)_yFe_{2-z}Se_2$ with various nominal values of $x$ synthesized at the intercalation temperature, $T_i$, = 45 ˚C and 100 ˚C.

| $T_i$ (˚C) | Li content ($x$) | Mass loss (%) | HMDA content ($y$) |
|---|---|---|---|
| 45 | 0.25 | 9.53 | 0.25 |
| 45 | 0.5 | 9.14 | 0.24 |
| 45 | 1 | 19.39 | 0.57 |
| 45 | 2 | 32.55 | 1.18 |
| 45 | 3 | 25.24 | 0.84 |
| 45 | 4 | 22.42 | 0.74 |
| 100 | 0.25 | 7.70 | 0.19 |
| 100 | 0.5 | 15.13 | 0.42 |
| 100 | 1 | 23.16 | 0.72 |
| 100 | 1.25 | 27.24 | 0.89 |
| 100 | 1.5 | 25.03 | 0.80 |
| 100 | 2 | 27.64 | 0.93 |